\documentclass[twocolumn, showpacs, preprintnumbers, amsmath, amssymb]{revtex4}
\usepackage{dcolumn}
\usepackage{graphicx}
\usepackage{bm}
\begin{document}
%--------------------------------------------------------------
\title{Networks between Professionals and Society: A Model for
Protein Dependency}
%--------------------------------------------------------------
\author{Henning Frydenlund Hansen}
\email{Henning.Hansen@ntnu.no}
\author{Alex Hansen}
\email{Alex.Hansen@ntnu.no}
\affiliation{Department of Physics, Norwegian University of Science and
Technology, N--7491 Trondheim, Norway}
%--------------------------------------------------------------
\date{\today}
%--------------------------------------------------------------
\begin{abstract}
We propose a network model with a fixed number of nodes and links with a dynamics which favors 
links between nodes differing in connectivity. Parameter regimes where the degree 
distributions follow power-laws, $P(k)\sim k^{-\gamma}$, high clustering following 
$C(N)\sim 1/N$ and small-world properties, with a network diameter following 
$D(N)\sim A+ B\log N$, are observed. Our model gives results comparable with real-world 
protein networks. 
\end{abstract}
%--------------------------------------------------------------
\maketitle
%--------------------------------------------------------------
\section{Introduction}

Over the last few years a large number of network models have been put 
forward, highly motivated by empirical studies of real-world networks.

The various models can be categorized belonging to one of three main 
classes of modeling paradigms. First, different variants 
of the random graph model of Erd\H{o}s and R\'{e}nyi \cite{er60} 
are still used for comparison with many different models and empirical 
studies \cite{ab02}. The second group of network models are 
refered to as small-world models, first presented by Watts and Strogatz 
\cite{ws98} and are motivated by high clustering observed in 
many real-world networks. This group of network models aims to include 
both the idea of highly clustered networks and random graphs. 
Third, the construction of various scale-free models have been 
motivated by the discovering of power-law degree distributions in real-world 
networks, ranging from the World Wide Web \cite{ajb99} 
to the network of Science collaboration \cite{n01} and the web of 
human sexual contacts \cite{lensa01}. This group of network models 
focuses on the dynamics of the network and aims to offer a universal 
theory of network evolution \cite{ab02}.

In the past few years, a wide range of concepts and measures for complex 
networks have been proposed and investigated. However, complex networks are  
most often described by three basic concepts.

The small-world concept describes the fact that there is a relative short 
path between any two nodes in most networks. The maximum of the shortest 
paths between any two nodes, refered to as the diameter, is often observed 
to grow logarithmically with the network size, $N$. This property is not 
related to a particular organizing principle \cite{ab02}, 
and are observed in random graphs, small-world model networks and scale-free 
networks. 

The clustering of a network is related to the formation of cliques of nodes 
being linked to each other. The clustering in most real networks is observed 
to be larger than the clustering in random graphs. Many proposed models of 
complex networks grasp this idea.

The third main characteristic for complex networks is the degree distribution. 
The degree distribution, $P(k)$, gives the probability for a randomly selected 
node to be connected to $k$ different other nodes. For a wide range of 
complex networks a power-law distribution $P(k)\sim k^{-\gamma}$ has been 
observed \cite{ajb99,n01,lensa01,ab00,asbs00,r98,ms02}. This deviates 
significantly from random graphs where links are placed randomly and from 
small-world models. In random graph models and in small-world models a 
large number of nodes have a degree close to the average degree of the 
network, $\overline{k}$.

Over the last few years a wide range of protein networks have been 
studied \cite{ajb00,j00,j01,ms02}. These networks are formed by direct physical
interaction between pairs of proteins and they form the underlying structure 
for the propagation of various signals regulating the proteins \cite{ms02}.
The motivation for the present study is the recent observation that 
protein networks guiding the biochemistry of living cells, the placement
of links tends to occur between high and low connectivity nodes rather than
between nodes of similar connectivity \cite{ms02}. The fact that one observe 
that the highly connected proteins are mostly connected to those with low 
connectivity, meaning that the highly connected nodes are well separated, is
believed to increase the robustness of the networks \cite{ms02}. 
This observation is somewhat reminiscent of the networks that connect
people with respect to their professional specialities. An example 
might be a network describing medical relations in terms of 
physician-patients relations. A physician (a highly connected node) has 
many patients (low connected node), but does not have a physician-patient 
realtion to many other physicians. Patients do not have physician-patient 
relations with each other.  The same argument goes for many networks 
describing relations based on a certain profession or some sort of 
specialization.

In the next section, we describe how we construct networks with the property
that contrasting nodes are preferably attached through links by using a 
Monte Carlo technique.  We then go on to show the results of our model simulations, 
focusing on the three main characteristics of complex networks, degree distributions, 
clustering and small-world properties. Finally we construct randomized versions of our 
networks and calculate the ratio between the degree distributions from the original network
and the randomized version of the same network. These results are compared with results 
from real-world protein networks.

%--------------------------------------------------------------
\section{Description of the model}

Rather than attempting to construct a network that preferably connects 
highly contrasting nodes through reconstructing the process that
may naturally have developed them, we take the Monte Carlo approach.  Even
though this is well-known and well understood technique, it is 
useful to remind the reader of its philosophy as we proceed.  Given a 
probability distribution $p(i,j)$ for having a link between nodes $i$ and 
$j$, the Monte Carlo method constructs a biased random walk through the set 
of different network configuration such that the relative frequency of 
encountering configurations with such a link present is proportional to 
$p(i,j)$. This probability should not set any intrinsic scale for the
network. The natural candidate for such a probability should then
be a power law in the ratio $k_i/k_j$.  Hence, we 
propose the simplest form that
accentuates the contrast between the connectivity of the two nodes $i$ and $j$
without bringing any intrinsic scales into the problem,
\begin{equation}
\label{pij}
p(i,j)=
\left(\frac{\max\left( k_i, k_j \right)}{\min\left( k_i, k_j \right)}
\right)^\beta\;.
\end{equation}  
This may be rewritten in a more compact form as follows,
\begin{equation}
\label{pijexp}
p(i,j)=e^{-\beta H(i,j)}\;,
\end{equation}
where
\begin{equation}
\label{hij}
H(i,j)=-\left|\ln\left(\frac{k_i}{k_j}\right)\right|\;.
\end{equation}

The network we consider consists of $N$ nodes with $L$ undirected links
between them. We term a given configuration of links between the nodes as
$\mathcal{G}$.  The probability to find a given configuration $\mathcal{G}$
is then
\begin{equation}
\label{pg}
\mathcal{P}(\mathcal{G})=\prod_{links} p(i,j)=e^{-\beta H}\;,
\end{equation}
where
\begin{equation}
\label{hall}
H= \sum_{links}-\left|\ln\frac{k_i}{k_j}\right|\;.
\end{equation}
We see that formally, the probability for finding a given configuration
$\mathcal{G}$ follows the Boltzmann distribution with a Hamiltonian
defined in Eq.\ (\ref{hall}).  The parameter $\beta$ has the formal
appearence of an inverse temperature, but it should not be interpreted as
anything more than the single remaining parameter in the probability when 
scale-freeness is implemented.  

We implement the Monte Carlo procedure using the Metropolis algorithm
\cite{mrrtt53,v01}, a well-known algorithm also previously used in different 
network models \cite{bm03}. In order to construct the 
random walk, we need the transitional
probabilities $\mathcal{P}(\mathcal{G} \to \mathcal{G}')$ 
which have to obey detailed balance, $\mathcal{P}(\mathcal{G})\mathcal{P}
(\mathcal{G} \to \mathcal{G}')=\mathcal{P}(\mathcal{G}')\mathcal{P}
(\mathcal{G}' \to \mathcal{G})$. The Metropolis prescription consists in first
defining a set of neighborhood configurations.  These are in our case simply
all configurations that can be reached from $\mathcal{G}$ moving {\it one\/}
link without placing two links between the same pair of nodes.  
The number of such neighboring states is $\mathcal{L}=[N(N-1)/2]!
/([N(N-1)/2-L]!L!)$.  Next we define the partial transitional probability
$\pi(\mathcal{G} \to \mathcal{G}') = 1/\mathcal{L}$.  If we now have that
\begin{equation}
\label{detbal}
\pi(\mathcal{G} \to \mathcal{G}')=\pi(\mathcal{G}' \to \mathcal{G})=
\frac{1}{\mathcal{L}}\;,
\end{equation}
the Metropolis construction of $\mathcal{P}(\mathcal{G} \to \mathcal{G}')$ from
$\pi(\mathcal{G} \to \mathcal{G}')$ ensures that detailed balance is fulfilled.
By construction, relation (\ref{detbal}) is fulfilled.  The transitional
probability is now given by
\begin{align}
\label{transmet}
\mathcal{P}(\mathcal{G} \to \mathcal{G}')&=\pi(\mathcal{G} \to \mathcal{G}')
\min\left(1,\frac{\mathcal{P}(\mathcal{G}')}{\mathcal{P}(\mathcal{G})}\right) \nonumber \\
&=\frac{\min\left(1,e^{-\beta\Delta H}\right)}{\mathcal{L}}\;,
\end{align}
where 
\begin{equation}
\label{deltah}
\Delta H = H(\mathcal{G}')-H(\mathcal{G})\;.
\end{equation}
We emphasize at this point that it is essential that detailed balance to
be fulfilled if the Monte Carlo algorithm is to produce configurations
$\mathcal{G}$ with probability proportional with the prescribed
$\mathcal{P}(\mathcal{G})$ --- and that this is fully ensured once 
Eq.\ (\ref{detbal}) is fulfilled. 

%--------------------------------------------------------------
\begin{figure*}[ht!]
\begin{center}
\resizebox{\textwidth}{!}{\includegraphics{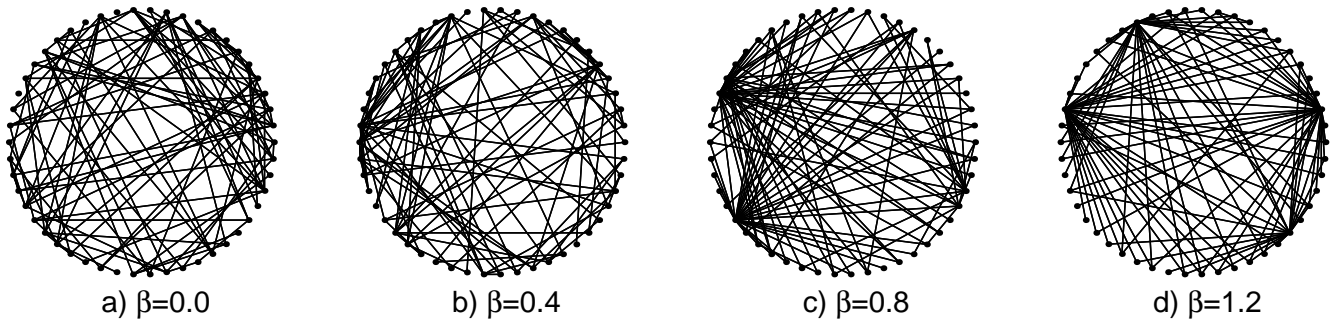}}
\caption{A network with $N=50$ nodes with $\overline{k}=4$ for different
inverse temperature $\beta$ when the networks have reached equilibrium 
after $\sim N^2$ iterations.}
\label{fig:ill}
\end{center}
\end{figure*}
%--------------------------------------------------------------

%--------------------------------------------------------------
\section{Results}

In figure \ref{fig:ill} we show examples of networks with $N=50$ nodes and
average connectivity $\overline{k}=4$ for different values of the parameter
$\beta$ when the networks have reached equilibrium after $\sim N^2$ iterations. 
Our model allows so-called single nodes, i.e., nodes that are not connected 
to any other nodes. It is also possible to split the network in disjoint 
components. 

\begin{figure}[h!]
\resizebox{\columnwidth}{!}{\includegraphics{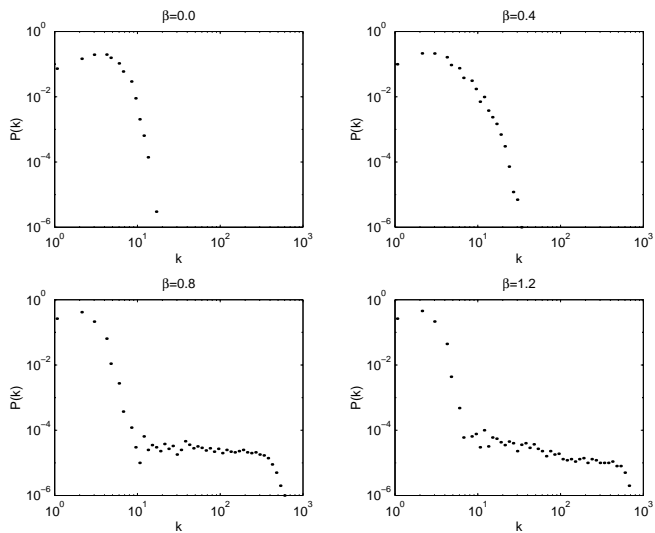}}
\caption{The probability $P(k)$ for a node to be linked to $k$ 
different other nodes for different $\beta$. 
These nodal distributions are averages for networks with $N=2000$ 
nodes with $\overline{k}=4$.}
\label{fig:con}
\end{figure}

In figure \ref{fig:con} we show the probability, $P(k)$, for a node
being linked to $k$ different other nodes for different values of the
parameter $\beta$.  For small values of $\beta$, the network behaves 
essentially as a random network. However, as $\beta$ is increased, the
importance of the contrast between the connectivity of each pair of nodes
is increasingly accentuated.  The effect of this is seen clearly in the
diagrams for $\beta \le 0.8$, where a power law appears.  This is the
regime where the model produces networks with the connectivity properties
described in the Introduction.  Furthermore, these networks are 
{\it scale free,\/} as power law in $P(k)$ indicates. 

\begin{figure}[h!]
\resizebox{\columnwidth}{!}{\includegraphics{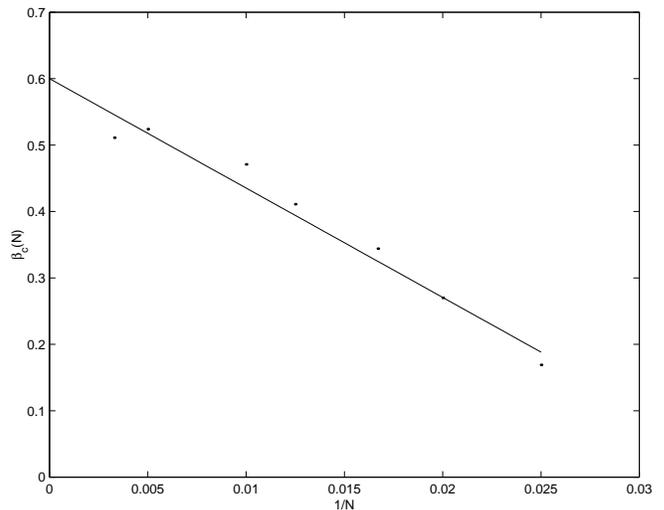}}
\caption{Slope intersection with the $\beta$-axis for the stepfunction 
$\sum_{k}k^{n} P_{N,\beta}(k) \,\, (n \ge 2)$ as function of 
inverse network size, $1/N$. Extrapolation gives a critical 
$\beta_{c}=0.60$.}
\label{fig:critical}
\end{figure}

There is a phase transition in the model associated with a $\beta=\beta_c$.
In order to investigate this, we study moments of the nodal distribution
$P_{N, \beta}(k)$, $\sum_{k}k^{n} P_{N,\beta}(k) \,\, (n \ge 2)$, for 
different network sizes $N$. As $N$ grows, 
$\sum_{k}k^{n} P_{N,\beta}(k) \,\, (n \ge 2)$ plotted as a function of 
$\beta$, converges towards a stepfunction with the step at critical $\beta_c$. 
By looking at the slope of the step and plot the slope intersection with 
the $\beta$-axis versus $1/N$ and finally extrapolating $1/N \rightarrow 0$, 
the numerical value of $\beta_c$ may be determined.  The result of this
analysis is shown in figure \ref{fig:critical}, and we find $\beta_c=0.60$.
 
%--------------------------------------------------------------
%--------------------------------------------------------------

%--------------------------------------------------------------
%--------------------------------------------------------------

\begin{figure}[h!]
\resizebox{\columnwidth}{!}{\includegraphics{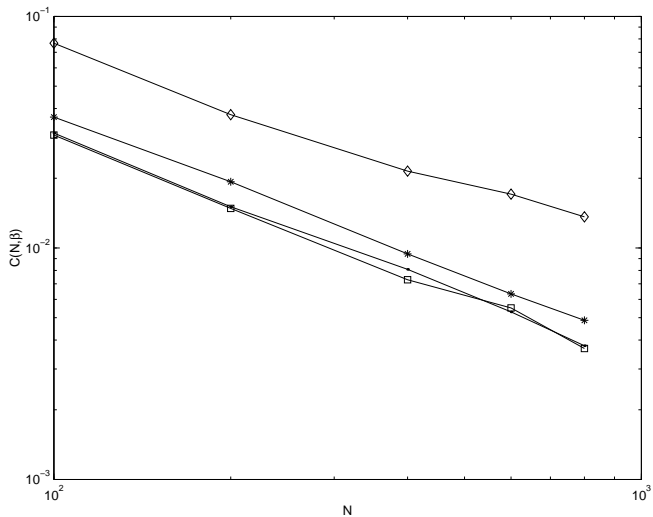}}
\caption{Mean cluster coefficient, $C(N,\beta)$ as a function of network 
size for different $\beta$ values. $\bullet: \beta=0.0$, 
$\star: \beta=0.4$, $\diamond: \beta=0.8$ and $\Box: \beta=1.2$.}
\label{fig:cluster}
\end{figure}

Figure \ref{fig:cluster} shows the cluster coefficient for
different values of the parameter $\beta$ as a function of the network 
size, $N$. If we consider a single node $i$ with connectivity $k_i$, 
which means $k_i$ neighbors, we calculate the cluster coefficient $c_i$ as i
$c_i=m_i/M_{k_i}$, where $M_{k_i}$ is the highest possible number of links i
between $k_i$'s neighbors, $M_{k_i}= \frac{k_i(k_i-1)}{2}$, while $m_i$ is 
the actual number of links between $k_i$'s neighbors. The mean cluster 
coefficient for a given temperature and a given network size, $C(\beta,N)$ 
is the average of all these $c_i$'s. For all values of $\beta$, we have a 
decreasing cluster coefficient as a function of the network size $N$, 
$C(\beta,N) \sim N^{-\omega}$ with $\omega$ close to $1$ for all $\beta$. 
We observe that the largest clustering is found for intermediate values of
$\beta$, close to $\beta_c$

%--------------------------------------------------------------
%--------------------------------------------------------------

\begin{figure}[h!]
\resizebox{\columnwidth}{!}{\includegraphics{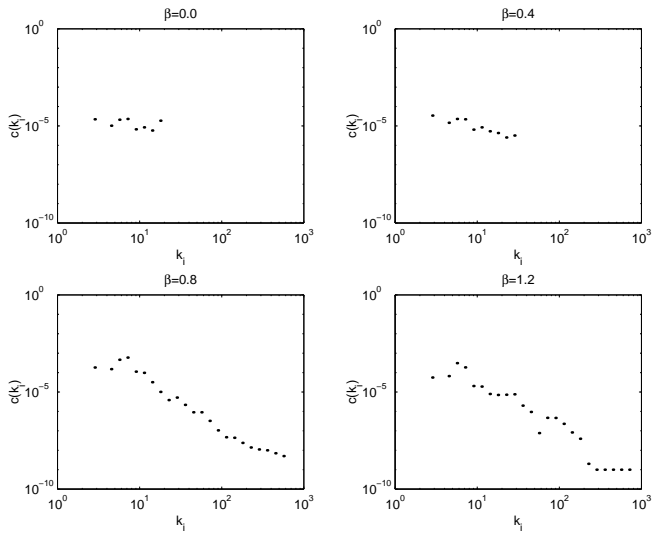}}
\caption{Cluster coefficient as a function of the degree, $k$, for 
different $\beta$ values.}
\label{fig:cluster_coeff}
\end{figure}

It is also possible to look at the average cluster coefficient 
for a node with connectivity $k_i$. These results are shown in figure 
\ref{fig:cluster_coeff} for different values of $\beta$. 
For high $\beta$, the model seems to show a power-law dependence for the 
clustering as a function of the degree, $k$, $c(k_i)\sim k^{-\gamma}$, 
with $\gamma$ close to $3$. This means that nodes with low connectivity 
are typically better clustered than nodes with high connectivity. 
For $\beta=0.0$, we have an exponent $\gamma=0$ and this seems to be close 
to the situation for any $\beta$ below the critical value.

%--------------------------------------------------------------
%--------------------------------------------------------------

\begin{figure}[h!]
\resizebox{\columnwidth}{!}{\includegraphics{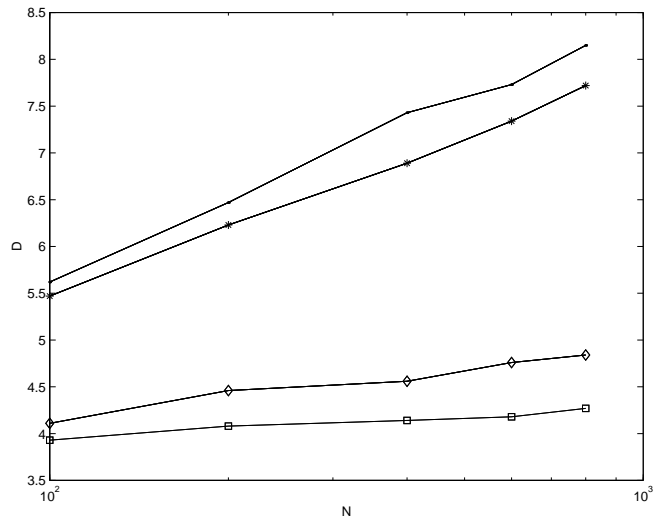}}
\caption{Mean diameter, $D(N)$, as a function of the network size, $N$, 
for different $\beta$. $\bullet: \beta=0.0$, $\star: \beta=0.4$, 
$\diamond: \beta=0.8$ and $\Box: \beta=1.2$.}
\label{fig:diameter}
\end{figure}

In many real-world networks one observes that there is a relative short 
path between any two nodes in the network. The maximum of the 
shortest paths between any two nodes in a network is most often refered 
to as the diameter of the network. 
In figure \ref{fig:diameter} we plot the mean diameter $D(N)$ as a 
function of the network size, $N$, for different $\beta$. We observe 
small-world properties with the diameter growing logarithmically 
with the network size, $N$.  Typically, we also see a growing diameter 
for lower $\beta$.

%--------------------------------------------------------------
%--------------------------------------------------------------

\begin{figure}[h!]
\resizebox{\columnwidth}{!}{\includegraphics{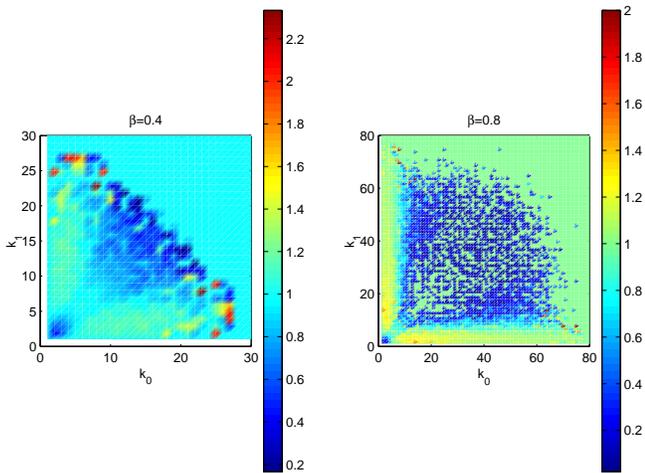}}
\caption{Connectivity correlations. $P(k_{0},k_{1})/P_{r}(k_{0},k_{1})$ 
where $P(k_{0},k_{1})$ is the probability for two nodes with connectivities 
$k_{0}$ and $k_{1}$ are connected, while $P_{r}(k_{0},k_{1})$ is the 
same probability in a randomized version of the same networks. 
In the randomized network, the nodes have the same connectivity as 
in the original network, but the links have been randomized.}
\label{fig:sneppen}
\end{figure}

A good illustration of connectivity correlations is to 
compare the network to a randomized network where the nodes have the same 
connectivity as in the original network but with randomized links. 
If $P(k_0, k_1)$ denotes the probability 
for a node with connectivity $k_0$ to be linked to a node with 
connectivity $k_1$ and $P_r(k_0,k_1)$ denotes the same 
probability in the randomized network, then the ratio 
$P(k_0,k_1)/P_r(k_0,k_1)$ is an interesting measure for the connectivity 
correlations which can extract some characteristics of the network. 
Figure \ref{fig:sneppen} shows this ratio for two values of $\beta$, on either 
side of the critical $\beta_c$, $\beta=0.4$ and $\beta=0.8$. Because our 
network consists of undirected links, we have a symmetry around $k_0=k_1$. 
We see regions in the $k_{0}-k_{1}$-plane where connections between nodes 
with certain connectivities either are significantly enhanced or suppressed 
compared to randomized networks. The darker blue area around $k_{0}=k_{1}$, 
reflects the tendency that it is less likely for two nodes with connectivities 
not differing much, to be connected. Along the $k_{0}$- or $k_{1}$-axis 
and close to these axis, we see a increased probability 
that nodes differing significantly in connectivity are connected.
The same ratio for real-world protein networks show richer patterns 
\cite{ms02}, but it is possible to recognize some tendencies comparing 
it to our results. In areas around $k_{0}=k_{1}$ we observe a reduced 
probability for two nodes with equal or close to equal connectivity 
to be connected, and we see an increased probability for nodes 
differing much in connectivity to be connected, observed close to the 
$k_0$- and $k_1$-axis \cite{ms02}.

%--------------------------------------------------------------
%--------------------------------------------------------------

%--------------------------------------------------------------
\section{Summary and Conclusion}

In this paper we have presented a network model with a static number of 
nodes and a static number of undirected links. Our model favors links 
between nodes differing in connectivity. We observe a series of 
characteristics observed in real-world networks. For small values of $\beta$,
the network behaves essencially as a random network. As $\beta$ increases,
the importance of the contrast between the connectivity of each pair of nodes
increases, and we observe scale free degree distributions indicated by power
laws. We also observe a phase transition at $\beta=\beta_c=0.60$. Our model 
gives networks with relative high clustering and a cluster coefficient
decreasing with increasing network sizes as $C(N)\sim N^{-1}$. We observe
the largest clustering for intermediate values of $\beta$ close to $\beta_c$.
The small-world property in our network model is indicated by a diameter growing 
logarithmically with the network size, $D(N)\sim A + B\log N$. The diameter
decreases for increasing values of $\beta$. Finally we constructed randomized 
versions of our networks in order to compare our results with real-world protein
networks. Patterns in the connectivity correlation plot, $P(k_o,k_1)/P_r(k_0,k_1)$, 
for $\beta$-values on either side of the critical value for $\beta$, show similarities
with patterns from real-world protein network. Connections between nodes with connectivities
differing significantly are enhanced while connections between nodes with equal 
or almost equal connectivities are suppressed compared to randomized versions of the same networks.  

We would like to thank Kim Sneppen at NORDITA for stimulating and fruitful discussions.

%--------------------------------------------------------------

%--------------------------------------------------------------

\end{document}